\documentclass[]{pasj01}
\let\leftroot\relax
\let\uproot\relax

\let\leftrightarrows\relax
\let\gtrsim\relax

\let\lesssim\relax

\let\varnothing\relax
\usepackage[switch,mathlines]{lineno}
\usepackage{natbib}
\usepackage{amsmath}
\usepackage{amssymb}
\makeatletter
\@ifpackageloaded{hyperref}{}{\newcommand{\href}[1]{}}
\makeatother
\usepackage{subfigure}
\usepackage{color}
\twocolumn

\Received{}
\Accepted{}
 
\begin{document} 

\title{First star formation in extremely early epochs}

\author{Mana \textsc{Ito} \altaffilmark{1}}
\altaffiltext{1}{Astronomical Institute, Graduate School of Science, Tohoku University, Aoba, Sendai 980-8578, Japan}
\email{mana.ito@astr.tohoku.ac.jp}

\author{Kazuyuki \textsc{Omukai} \altaffilmark{1}}
\email{omukai@astr.tohoku.ac.jp}


\KeyWords{stars: formation --- stars: Population III --- cosmic background radiation --- dark ages, reionization, first stars}

\maketitle

\begin{abstract}

First stars play crucial roles in development of the universe, influencing events like cosmic reionization and the chemical enrichment of the intergalactic medium. While first stars are conventionally thought to form at around $z \sim 20-30$ in the standard $\Lambda$ Cold Dark Matter ($\Lambda$CDM) cosmology, observational constraints on small-scale ($<$ Mpc) density fluctuations remain limited, possibly differing significantly from the scale-invariant fluctuations assumed in the $\Lambda$CDM model. Should this be the case, the formation of first stars could occur much earlier than typically predicted.
In this study, we investigate the formation process of first stars in the extremely early epochs of $z \gtrsim 100$ in the post-recombination universe. At such early times, the effects of the warm cosmic microwave background (CMB) become significant. We calculate the collapse of primordial star-forming clouds using a one-zone thermo-chemical model that accounts for CMB influences on radiative heating, Compton cooling, and photodissociation reactions.
We found that the impact of the CMB on the evolution is limited at $z \lesssim 100$, with the temperature evolution closely resembling the conventional model. However, within the range $100 \lesssim z \lesssim 400$, the formation of H$_{2}$ via the H$^{-}$ channel is impeded by H$^{-}$ photodetachment induced by the CMB, leading to higher temperatures compared to standard thermal evolution. Consequently, first stars with masses exceeding $1000 ~\mathrm{M}_{\odot}$ can emerge at $z \gtrsim 100$. Furthermore, at $z \gtrsim 500$, the temperature evolution becomes nearly isothermal at several thousand Kelvins solely due to atomic cooling, as H$_{2}$ formation is entirely suppressed, including the less efficient H$_2^{+}$ channel, which is blocked by H$_2^{+}$ photodissociation.
In such cases, supermassive stars with masses around $\sim 10^{5} ~\mathrm{M}_{\odot}$ are expected to form solely via atomic cooling. These findings emphasize the significant variation in the typical mass of the first stars depending on the epoch of formation.
\end{abstract}


\section{Introduction}
The launch of the James Webb Space Telescope (JWST) has revolutionized our exploration of the high-redshift universe, shedding new light on the cosmic dawn. Early observations with the JWST have unveiled a substantial population of bright galaxies at redshifts greater than 10 \citep{Naidu2022, Harikane2023, Donnan2023}. While some galaxies initially identified through photometric observations have been revealed as low-redshift interlopers upon subsequent spectroscopic analysis, over ten galaxies have been spectroscopically confirmed at redshifts exceeding 10 {\citep{ArrabalHaro2023a, ArrabalHaro2023b, Curtis-Lake2023, Fujimoto2023, Harikane2024} }. Their unexpectedly high number density, approximately an order of magnitude greater than extrapolations from lower redshifts, has introduced tension with models assuming a constant star formation efficiency \citep{Harikane2024}. Additionally, there are candidates for extremely massive galaxies ($M_{\ast} \sim 10^{10-11}~\mathrm{M}_{\odot}$) at somewhat lower redshifts, around $z \sim 7-9$ \citep{Labbe2023}. The presence of these galaxies challenges the standard $\Lambda$CDM scenario, suggesting either exceptionally high star formation efficiencies or the need for entirely new theoretical frameworks \citep{Inayoshi2022, Boylan-Kolchin2023, Lovell2023}.

These findings have prompted the proposal of various astrophysical mechanisms, aimed at explaining the abundance of bright galaxies at high redshifts. These include suggestions of a top-heavy initial mass function (IMF) in early galaxies \citep{Chon2021, Chon2022, Chon2023}, high star formation efficiencies \citep{Fukushima2021, Dekel2023}, or reduced dust extinction \citep{Ferrara2023}. Ongoing research actively investigates the feasibility of these proposals.
 
On the other hand, alternative scenarios have been proposed that go beyond the framework of the standard $\Lambda$CDM cosmology. The cosmological structure formation process involves density fluctuations present in the early universe growing under gravity, becoming nonlinear, and collapsing gravitationally to form dark matter halos, within which stars form. The scales at which halos form at different epochs are determined by the spectrum of primordial fluctuations. Since the amplitude of primordial fluctuations on scales smaller than 1 Mpc has not been directly observed, it is typically assumed that scale-invariant fluctuations on larger scales continue on smaller scales as well \citep[see discussions in, e.g.,][]{Kohri2008, Bringmann2012, Byrnes2019}. However, the possibility of larger fluctuations existing on smaller scales cannot be ruled out. If such fluctuations did exist, it would imply the formation of galaxies at earlier cosmic epochs.

For instance, \cite{Padmanabhan2023} and \cite{Tkachev2024} have proposed models featuring a bumpy enhancement in the spectrum of fluctuations on sub-Mpc scales. Similarly, \cite{Parashari2023} and \cite{Hirano2024} have considered a blue-tilted spectrum with a bend on the small-scale side. In the former works of each pair, the number of galaxies formed was estimated analytically, while the latter conducted $N$-body numerical simulations. Both studies argue that the presence of bright high-redshift galaxies observed by JWST can be explained by these models.

Various factors contribute to fluctuations on small scales, including the running of the power spectrum arising from certain inflation models \citep[e.g., ][]{Kohri2008, Parashari2023} and phase transitions in the early universe \citep[e.g.,][]{Ricotti2009}. Recently, there has been increasing research interest in the possibility of primordial black holes (PBHs) forming in the early universe \citep[e.g.,][]{Carr2021, Green2021}. If PBHs exist, their spatial distribution is known to generate isocurvature fluctuations, leading to significant fluctuations on small scales. It has been suggested that if PBHs comprise a small fraction of dark matter ($\sim 10^{-6}-10^{-3}$), the observed number density of high-redshift galaxies could be explained by this effect \citep{Liu2022}.

Additionally, recent observations from Pulsar Timing Arrays, such as NanoGrav15yr \citep{Agazie2023}, suggest the presence of background gravitational waves . These could be interpreted as cosmological non-linear secondary gravitational waves, leading to large density fluctuations on small scales and the generation of PBHs \citep{Inomata2023}. There are suggestions that fluctuations may not be large enough to form PBHs, which might be more plausible, but even in such cases, non-linearities in the evolution of the universe could lead to the early formation of halos \citep{Ricotti2009, Abe2022}.
Furthermore, the presence of primordial magnetic fields has been proposed to generate secondary density fluctuations on small scales, potentially leading to earlier galaxy formation \citep{Tashiro2006, Adi2023}.

Due to the factors mentioned above, if the formation of first stars occurs much earlier than typically assumed, it is expected that the evolution of star-forming clouds and the resulting masses of the first stars will differ from the usual pattern.

In the standard $\Lambda$CDM scenario, the formation of first stars is conventionally hypothesized to occur within the redshift range of $z=20-30$ \citep{Couchman1986, Haiman1996, Tegmark1997}. Even in rare scenarios where only one such object might be observable in the universe, the formation epoch is estimated to be approximately $z=60-70$ \citep{Naoz2006, Fialkov2012}. Consequently, the formation of first stars beyond $z>100$ remains scarcely explored.

In this context, \cite{Hirano2015} conducted cosmological simulations to explore the process of first star formation in very early epochs ($z\simeq 100-200$), particularly in scenarios where the matter power spectrum exhibits a blue tilt on small scales. They found that 
the dissociation of H$^{-}$ by the high-temperature cosmic microwave background (CMB) inhibits the formation of H$_2$, a crucial coolant for primordial gas. Consequently, the star-forming gas becomes hotter than usual. They concluded that the first stars formed under such conditions tend to be somewhat more massive.


In this paper, we explore the possibility of first star formation occurring in the universe significantly earlier than what is typically assumed in the standard cosmological scenario. 

We extend the investigation of potential formation epochs for the first stars from the typical early star formation epoch (around a few tens of redshifts) to just after cosmological recombination (around a few hundred redshifts), extending beyond the scope of the study conducted by \cite{Hirano2015}, who calculated the evolution of several first-star-forming clouds forming within a redshift range of $z \sim 20-200$ selected from cosmological simulations. Here, we systematically study the thermal and chemical evolution of such clouds in a much wider redshift range of $z \sim 20-700$. In addition, we calculate the entire prestellar collapse evolution up to the protostellar density of $n_{\rm{H}}=10^{22}$ cm$^{-3}$, whereas \cite{Hirano2015} stopped their calculation at $n_{\rm{H}}=10^{13}$ cm$^{-3}$.

We aim to understand how the thermal evolution of star-forming clouds would differ across this broad range of formation epochs and discuss the properties of the stars formed under these conditions, using a one-zone thermo-chemical model.

The structure of this paper is as follows: Section 2 outlines the methodology, followed by Section 3, which presents the computational results. In Section 4, we explore the properties of the stars formed based on these computational results and discuss other relevant aspects.
Finally, in Section 5, we provide a brief summary.
For the cosmological parameters, we adopt the following values: $\Omega_{\Lambda} = 0.76$, $\Omega_{\rm m} = 0.24$, $\Omega_{\rm b} = 0.04$ and $H_{0} = 70 ~\mathrm{km ~s^{-1} ~Mpc^{-1}}$.

\section{Method of Calculation}

In this paper, we calculate the thermal and chemical evolution process of primordial clouds in high-density cosmological regions, 
starting from the initial expansion, maximum expansion, gravitational contraction, and virialization up to the formation of protostars after further gravitational collapse, using a one-zone model. We employ a modified version of the computational model used by \cite{Nakauchi2019}, who investigated the thermal evolution during the typical first star formation \citep[see also][]{Omukai2000, Omukai2005}. 
Here, we outline the model and describe the modifications made for the extension of calculations to extremely high-redshift periods.

\subsection{Pre-Virialization Evolution}

The primary focus of this calculation is the thermal evolution during the contraction of high-density cores, which significantly impacts the mass of forming stars. To establish initial conditions, it is necessary to compute the pre-virialization evolution.

Initially, for each halo's virialization epoch $z_{\rm vir}$, the density evolution is computed until reaching the virialization density using the spherical top-hat density evolution model \citep[e.g.,][]{Padmanabhan1993} as follows:
\begin{equation}
\rho_{\rm tot} = \frac{9}{2}\pi^2
\left( \frac{1+z_{\text{turn}}}{1-\cos\theta} \right)^3 \rho_{{\rm cr},0} \Omega_{m}
\end{equation}
where $\rho_{{\rm cr},0}$ is the current critical density of the universe ($=3H_{0}^2/8\pi G$), and $z_{\rm turn}$ is the turnaround redshift, which is related to $z_{\rm vir}$ as $1+z_{\rm turn}=2^{2/3} (1+z_{\rm vir})$. The parameter $\theta$ is related to the redshift as
\begin{equation}
1+z = \left(1+z_{\text{turn}}\right)\left(\frac{\theta-\sin\theta}{\pi}\right)^{-2/3}.
\end{equation}
Additionally, the total matter density 
$\rho_{\rm tot}$
comprises contributions from both dark matter ($\rho_{\rm DM}$) and baryons ($\rho$). It is assumed that prior to virialization, the amounts of dark matter and baryons are proportional to the cosmic mean.

The evolution of ionization and temperature until the maximum expansion of the halo ($z=z_{\rm turn}$) is not calculated using our model for this work. Instead, we utilize RecFast \citep{Seager1999}, which accurately computes the recombination processes in the early universe. However, during this computation, the density evolves not monotonically as in the original RecFast code, resembling the evolution dictated by the aforementioned spherical top-hat model corresponding to each $z_{\rm vir}$.

Subsequent to $z_{\rm turn}$, when calculating the evolution of baryonic gas as it contracts, the thermal evolution code is switched to the one used in this study, evolving until the density reaches the virialization density.

As outlined above, for clouds at each formation epoch $z_{\rm vir}$, the state at $z_{\rm turn}$, which serves as the initial condition for our thermal evolution calculations, is uniquely determined by the density as prescribed by Equation (1). Additionally, the temperature and ionization degree are obtained by solving the cosmic recombination thermal evolution up to the time $z_{\rm turn}$
using the RecFast code, based on the density evolution given by Equation (1).
RecFast does not account for molecular reactions such as those involving hydrogen molecules. Therefore, for the abundance of hydrogen molecules, we adopted the values presented as a function of 
$z$ in Figure 4 of Galli and Palla (1998), who solved detailed chemical reactions and temperature evolution in a uniform early universe intergalactic medium (IGM).
The abundances of other minor species were set to be zero. These assumptions regarding chemical abundances have no significant impact on our results.

\subsection{Post-Virialization Evolution}

Once the density of the high-density region under consideration reaches the virialization density $\rho_{\rm vir, DM} = 8 \rho_{\rm DM}(z_{\rm turn})$, where $\rho_{\rm DM}$ denotes the dark matter density at $z_{\rm turn}$, we consider the dark matter to have virialized, and its density remains constant at this value thereafter. Note that since $z_{\rm vir}$ corresponds to the time when the density diverges in the top-hat model, the time when the density reaches the virialization density is slightly earlier than this. On the other hand, the CMB temperature $T_{\rm CMB}$ varies with time until $z_{\rm vir}$, after which it is fixed to the value corresponding to each virialization epoch.

Meanwhile, the baryonic gas initiates a runaway collapse \citep{Larson1969, Penston1969}. Assuming nearly free fall, we increase its density according to the timescale $t_{\rm ff} = \sqrt{{3\pi}/{32G\rho_{\rm tot}}}$ as follows:
\begin{equation}
\frac{d\rho}{dt} = \frac{\rho}{t_{\rm ff}}.
\end{equation}
It should be noted that runaway collapse can only occur if the gas mass consistently exceeds the instantaneous Jeans mass throughout the protostellar collapse.

The temperature evolution of the gas is computed by solving the energy equation:
\[
\frac{de}{dt} = -P\frac{d}{dt}\left(\frac{1}{\rho}\right) - \Lambda_{\rm net},  
\]
where the pressure is given by $P = \rho k_B T/\mu m_H$, the specific internal energy by $e = P/(\gamma_{\rm ad} - 1)\rho$, and $\Lambda_{\rm net}$ represents the net cooling rate per unit mass. The other symbols retain their usual meanings.

Here, the considered cooling and heating processes for $\Lambda_{\rm net}$ include line emission from the primordial gas (such as H Ly$\alpha$, H$_2$, and HD) and continuum emission (mainly H$_{2}$ collision-induced emission (CIE) and H$^{-}$ free-bound emission), as well as heating and cooling processes associated with chemical reactions. Additionally, in this study, the influence of the early universe's CMB is considered, incorporating the Compton cooling process. The cooling rate for this process is provided by the formula from \cite{Omukai2001} (equation B23).

Furthermore, in \cite{Nakauchi2019}, the calculation of line cooling rates was conducted using fitting functions without solving for the occupation numbers of energy levels. In our case, to account for the effect of heating by the CMB, we replaced the cooling rate with $\Lambda(T)-\Lambda(T_{\rm CMB})$ during the computation.

The continuum cooling rate was computed in \cite{Nakauchi2019} by assuming local thermodynamic equilibrium (LTE) and utilizing the Planck mean opacity. This approach is a good approximation for typical first star formation scenarios, where only H$_2$ CIE is crucial for cooling among continuum processes. However, in our calculations for star formation at extremely high redshifts, where clouds can reach temperatures of several thousand degrees, there are opacity regimes not covered by available data, and chemical equilibrium might not hold as usual. Therefore, assuming local thermodynamic equilibrium for computing the cooling rate may no longer be a good approximation. To address this issue and consistently provide continuum cooling rates even in such cases, we computed the continuum cooling rate (Table A2) as well as Planck and Rosseland mean opacity (Tables A3 \& A4) using the method outlined in \cite{Matsukoba2019}.

The optical depth for continuum radiation is treated in the grey approximation. Taking into account scattering effects, it is computed as $\tau=\sqrt{\tau_{\rm R} \tau_{\rm P}}$ 
\citep[see e.g., Sec. 1.7 of][]{Rybicki1986}, where $\tau_{\rm R}$ and $\tau_{\rm P}$ represent the optical thickness of the cloud estimated using Rosseland and Planck mean opacity, respectively. It is worth noting that while Rosseland mean opacity considers both absorption and scattering, Planck mean opacity only accounts for pure absorption.

The size of the cloud is determined based on the following considerations. We model the star-forming cloud as uniform, corresponding to solving the evolution of the central region of the high-density core undergoing runaway collapse. The size of this central region is approximately the local Jeans length at that time, given by $\lambda_{\rm J} = \sqrt{\pi k_{\rm B} T/G\mu m_{\rm H} \rho_{\rm tot}}$, where $\rho_{\rm tot}$ denotes the instantaneous total density. When considering optical thickness, we estimate the cloud size as $\lambda_J$. Using the instantaneous gas density $\rho$, the optical thickness is given by $\tau_{\rm R, P} = \kappa_{\rm R, P} \rho \lambda_{\rm J}$.

The abundance of chemical species is determined by solving the chemical reactions of primordial gas with the temperature and density evolution computed as described above. Here, we consider a total of 23 species: H, D, He, Li, and their compounds, ${\mathrm{H}}$, ${\mathrm{H_2}}$, ${\mathrm{e^-}}$, ${\mathrm{H^+}}$, ${\mathrm{H_2^+}}$, ${\mathrm{H_3^+}}$, ${\mathrm{H^-}}$, ${\mathrm{He}}$, ${\mathrm{He^+}}$, ${\mathrm{He^{2+}}}$, ${\mathrm{HeH^+}}$, ${\mathrm{D}}$, ${\mathrm{HD}}$, ${\mathrm{D^+}}$, ${\mathrm{HD^+}}$, ${\mathrm{D^-}}$, ${\mathrm{Li}}$, ${\mathrm{LiH}}$, ${\mathrm{Li^+}}$, ${\mathrm{Li^-}}$, ${\mathrm{LiH^+}}$, ${\mathrm{Li^{2+}}}$, and ${\mathrm{Li^{3+}}}$, and we solve the reactions between them, as listed in Table 1 of \citet{Nakauchi2019}. All reactions, including forward and reverse reactions, are paired (total of 107 pairs), and the reverse reaction rates are calculated from the equilibrium constants using the forward reaction rates. Therefore, when the density becomes sufficiently high, the chemical composition is modeled to approach the correct chemical equilibrium values given by the Saha equations. Additionally, we include the photodissociation reaction of H$_2$ as described below.

In \cite{Nakauchi2019}, photochemical reactions were considered in cases when the cloud becomes optically thick and the radiation field within it is dominated by thermal radiation. 
The radiation intensity is given by
\begin{equation}
J_{\nu}= (1-e^{-\tau})B_{\nu}(T),
\end{equation}
where $B_{\nu}(T)$ is the Plack function
and the photochemical reaction coefficient is calculated as
\begin{equation}
k_{\rm dissoc}= (1-e^{-\tau})k_{\rm assoc} K_{\rm eq}(T),
\end{equation}
where $\tau$ represents the optical thickness of the cloud for continuum radiation, $k_{\rm assoc}$ is the reverse radiative association reaction coefficient and $K_{\rm eq}(T)$ is the equilibrium constant. 
However, in our case, since the strong CMB radiation is irradiating from outside, we incorporate the CMB into the radiation field as well:
\begin{equation}
J_{\nu}=(1-e^{-\tau})B_{\nu}(T)+ e^{-\tau}B_{\nu}(T_{\rm CMB}),
\end{equation}
and accordingly modify the photochemical reaction coefficient as:
\begin{eqnarray}
k_{\rm dissoc}&=&(1-e^{-\tau}) k_{\rm assoc}(T) K_{\rm eq}(T) \\
\nonumber
& & + e^{-\tau} k_{\rm assoc}(T_{\rm CMB}) K_{\rm eq}(T_{\rm CMB}).
\end{eqnarray}
In our case, particularly important photochemical reactions include the photodetachment of H$^-$ and photodissociation of H$_2^+$. The reaction rates for these species are treated as described above. Additionally, we incorporate the photodissociation of H$_2$, which was not considered in \citet{Nakauchi2019}. The reaction rate is determined using the expression from \citet {Sugimura2014}.
However, as we will discuss later, there were no instances where photodissociation of H$_2$ due to the CMB became significant.

\section{Results}

In this section, we present the results of the thermal and chemical evolution of primordial gas clouds as they collapse to form protostars at each virialization epoch $z_{\rm vir}=20-700$. We investigate the thermal processes underlying their collapse, with particular emphasis on the influence of the CMB at high redshifts.

\subsection{Early Cosmic Expansion Phase}
\begin{figure}[hbtp]
    \centering
	\includegraphics[keepaspectratio, bb=0 0 6144 4608, width = \columnwidth]{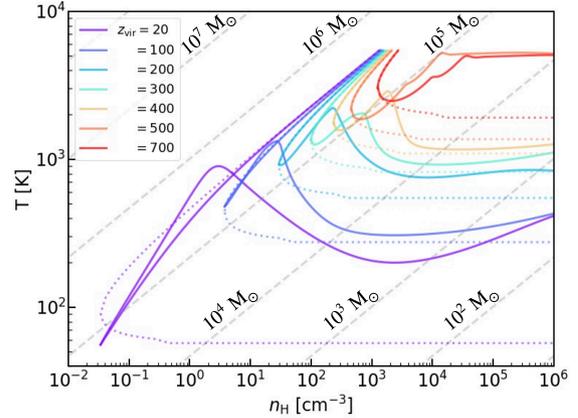}
	\caption{Early temperature evolution of primordial-gas star-forming clouds from maximum expansion to subsequent collapse
    for different formation epochs $z_{\rm vir}=20, 100, 200, 300, 400, 500$, and $700$. The temperature is plotted as a function of density, with different colors representing different formation epochs as indicated in the legend. The evolution starts at a temperature of approximately 6000K and a number density of about $10^3 ~\mathrm{cm^{-3}}$ at $z=2000$, then initially moves downward and to the left due to cosmic expansion, reaching maximum expansion before collapsing toward higher densities. Additionally, the figure includes dotted lines of the same color, indicating the CMB temperature at each density of the star-forming cloud. The grey dashed lines represent constant Jeans masses, with their values annotated in the figure.
}
	\label{Fig_nT_exp}
\end{figure}

First, we see the initial evolution of primordial gas clouds, starting with their expansion due to cosmic expansion, reaching maximum expansion, followed by contraction towards virialization, resulting in the onset of runaway collapse of cloud cores driven by self-gravity.

The temperature evolution of primordial clouds at such early phases is shown in Figure \ref{Fig_nT_exp} for different formation epochs $z_{\rm vir}=20, 100, 200, ..., 700$, as a function of number density. We also depict the CMB temperature when each cloud reaches its respective density by dashed lines.

Shown is the evolution from a pre-recombination epoch of $z=2000$ downward.  Each cloud experiences expansion initially, following the cosmic expansion, then reaches the maximum expansion before starting contraction. Note that both the virialization density, which is $18\pi^2$ times the mean density of the universe at that time, and the density at maximum expansion, which is 1/8 of the virialization density, are proportional to $(1+z_{\rm vir})^3$, reflecting the higher average density in the universe.	

At very early phases, the gas temperature decreases for the cases of any $z_{\mathrm{vir}}$. Throughout the early expansion phase, the gas temperature remains approximately equal to the CMB value for the cases with $z_{\rm vir} \gtrsim 100$, since the gas is tightly coupled to the CMB thermally via the Compton scattering. Consequently, the slope of the thermal evolution is shallower compared to that of adiabat (=2/3).
In cases of lower formation redshift, for example, $z_{\rm vir}=20$, corresponding to as in the standard epoch of Population III star formation, although initially at $T \gtrsim 600{\rm K}$, i.e., $z \gtrsim 200$, the gas is thermally coupled to the CMB tightly, at lower temperature and redshift, the gas thermally decoupled from the CMB, resulting in adiabatic expansion. 

For $z_{\rm vir} \gtrsim 200$, 
even after reaching the maximum expansion and after the onset of the collapse, the temperature continues decreasing further 
due to Compton coupling. Subsequently, with further compression, as heating due to gravitational compression exceeds the Compton cooling,
whose effect weakens also due to electron recombination in clouds, the temperature begins to rise with increasing density, eventually turning to adiabatic contraction.
At lower $z_{\rm vir}$, the temperature rises almost adiabatically from just after the maximum expansion.

Recall that the maximum Jeans mass during the contraction/virialization phases is a few $10^{5} M_{\odot}$ across all cases considered. This represents the minimum gas mass required for further collapse and subsequent star formation. Correspondingly, the required halo mass is $M_{\rm vir} = $ a few $10^{6} M_{\odot}$. 
It should be noted that we are considering haloes that are larger than this mass.
These haloes possess a compact virial radius, given by:
\begin{align}
r_{\rm vir} = \left[\frac{M_{\rm vir}}{\frac{4\pi}{3}18\pi^2 \rho_{\rm crit} \Omega_{\rm m} (1+z)^3 } \right]^{1/3} 
\label{eq:rvir}
\\
= 30  \text{pc} \left(\frac{M_{\rm vir}}{10^6 M_{\odot}} \right)^{1/3} \left(\frac{100}{1+z} \right),
\end{align}
where $\rho_{\rm crit}$ is the critical density of the universe, 
with dimensions similar to those of local molecular clouds.

\subsection{Thermal Evolution in Star-forming Cloud Cores}

Initially, baryonic gas and dark matter 
contract in the same way. After virialization of the halo, while the dark matter density becomes constant, the gas contracts further and its density continues increasing owing to energy dissipation due to radiative cooling as long as the halo is massive enough. 
The gravitational collapse of such high-density cloud cores eventually leads to the formation of protostars. Here, we examine the thermal evolution of star-forming clouds at these higher densities.

\begin{figure}[hbtp]
    \centering
	\includegraphics[keepaspectratio, bb=0 0 6144 3948, width = \columnwidth]{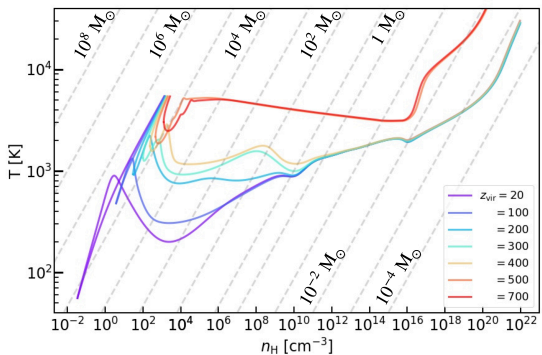}
	\caption{The temperature evolution of the same star-forming clouds as depicted in Figure 1, but focusing on higher densities up to the protostar formation epoch. The grey dashed lines represent constant Jeans masses. For lines corresponding to masses larger (smaller) than 1 $\mathrm{M}_{\odot}$, the gas is assumed to be in atomic (molecular, respectively) state.
 }
	\label{Fig_nT}
\end{figure}
\begin{figure}[hbtp]
    \centering
	\includegraphics[keepaspectratio, bb=0 0 864 648, width = \columnwidth]{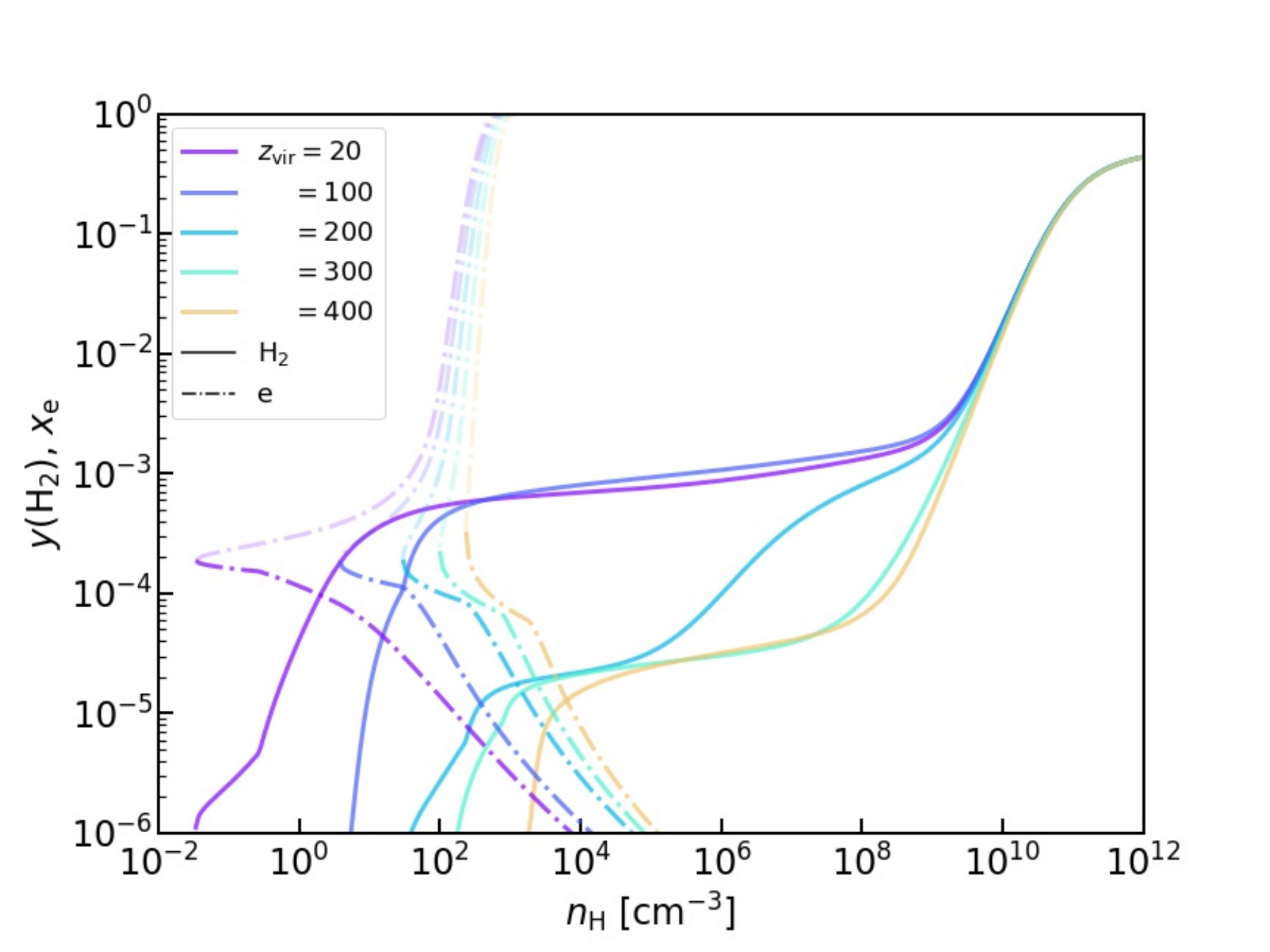}
	\caption{
 H$_{2}$ abundance (solid lines) and ionization fraction (dash-dotted lines) in primordial clouds collapsing at $z_{\rm vir}=20, 100, 200, 300, 400$. These clouds are capable of protostellar collapse due to H$_{2}$ cooling. For the ionization fraction lines, the lightly colored parts with larger values represent the values during the initial expansion phase.
 }
	\label{Fig_yH2}
\end{figure}
\begin{figure}[hbtp]
\centering
    \subfigure{%
	\includegraphics[keepaspectratio, bb=0 0 864 648, width = 0.95\columnwidth]{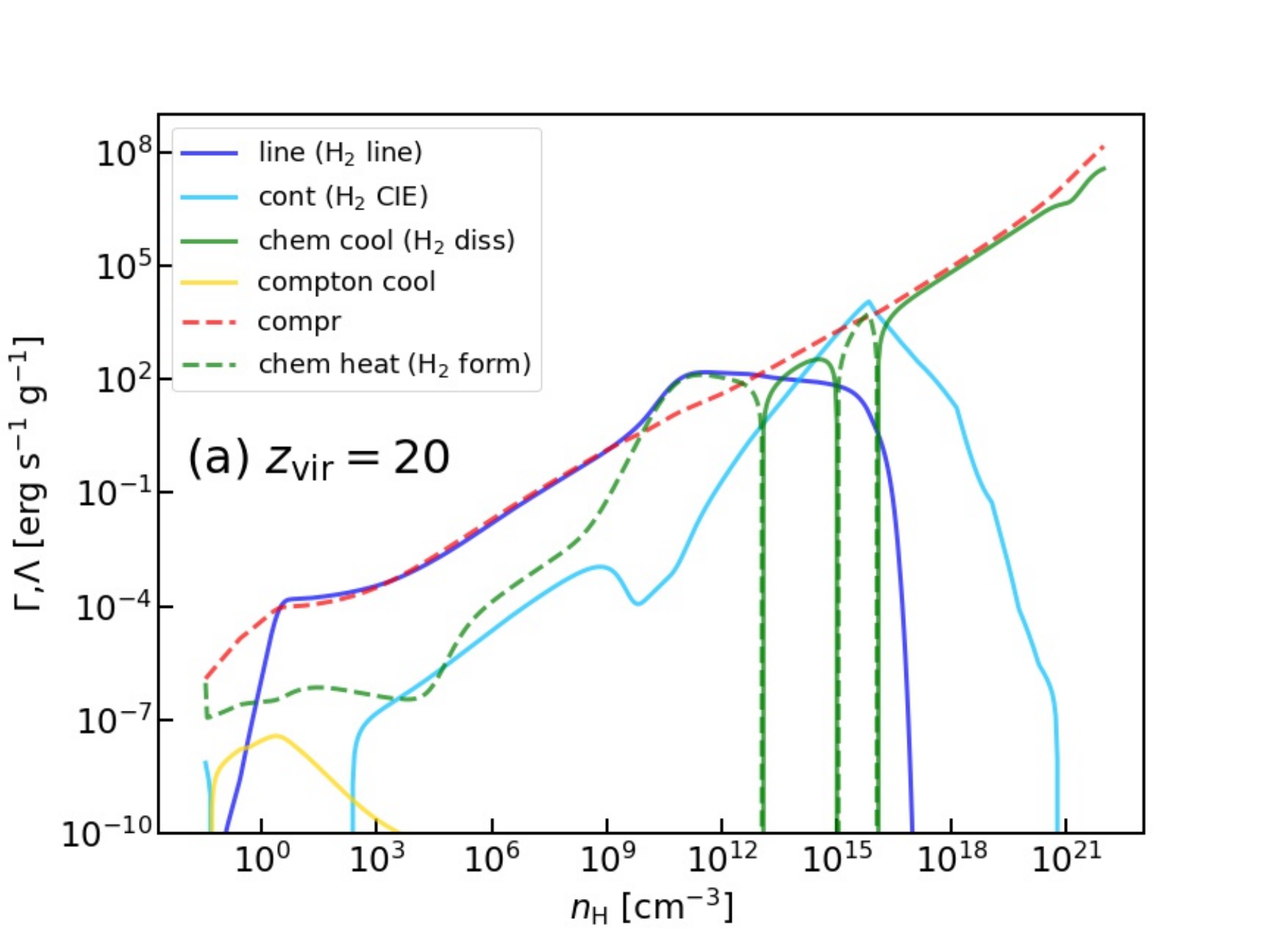}}\\%
    \vspace{0pt}
    \subfigure{%
	\includegraphics[keepaspectratio, bb=0 0 864 648, width = 0.95\columnwidth]{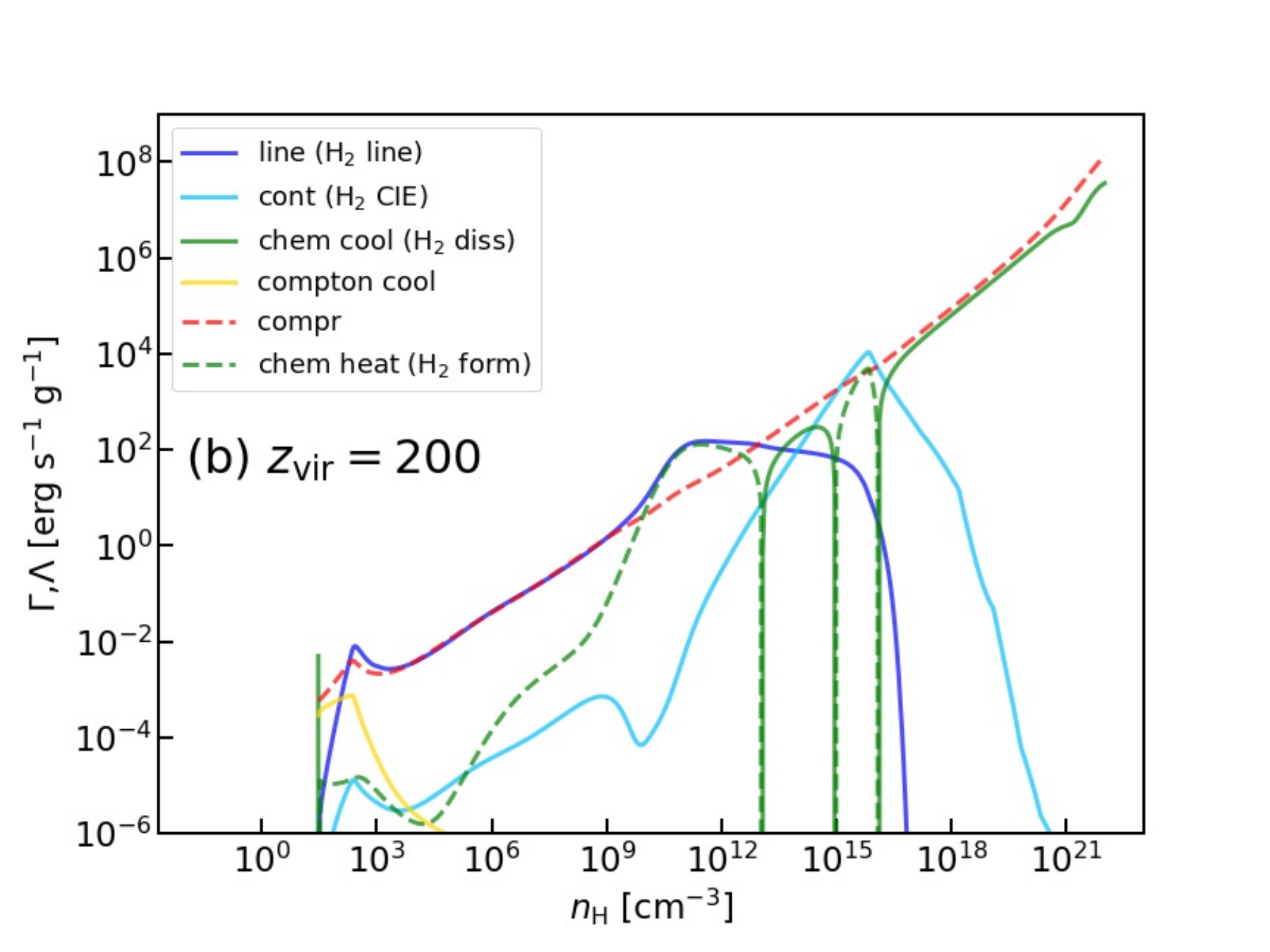}}\\%
    \vspace{0pt}
    \subfigure{%
	\includegraphics[keepaspectratio, bb=0 0 864 648, width = 0.95\columnwidth]{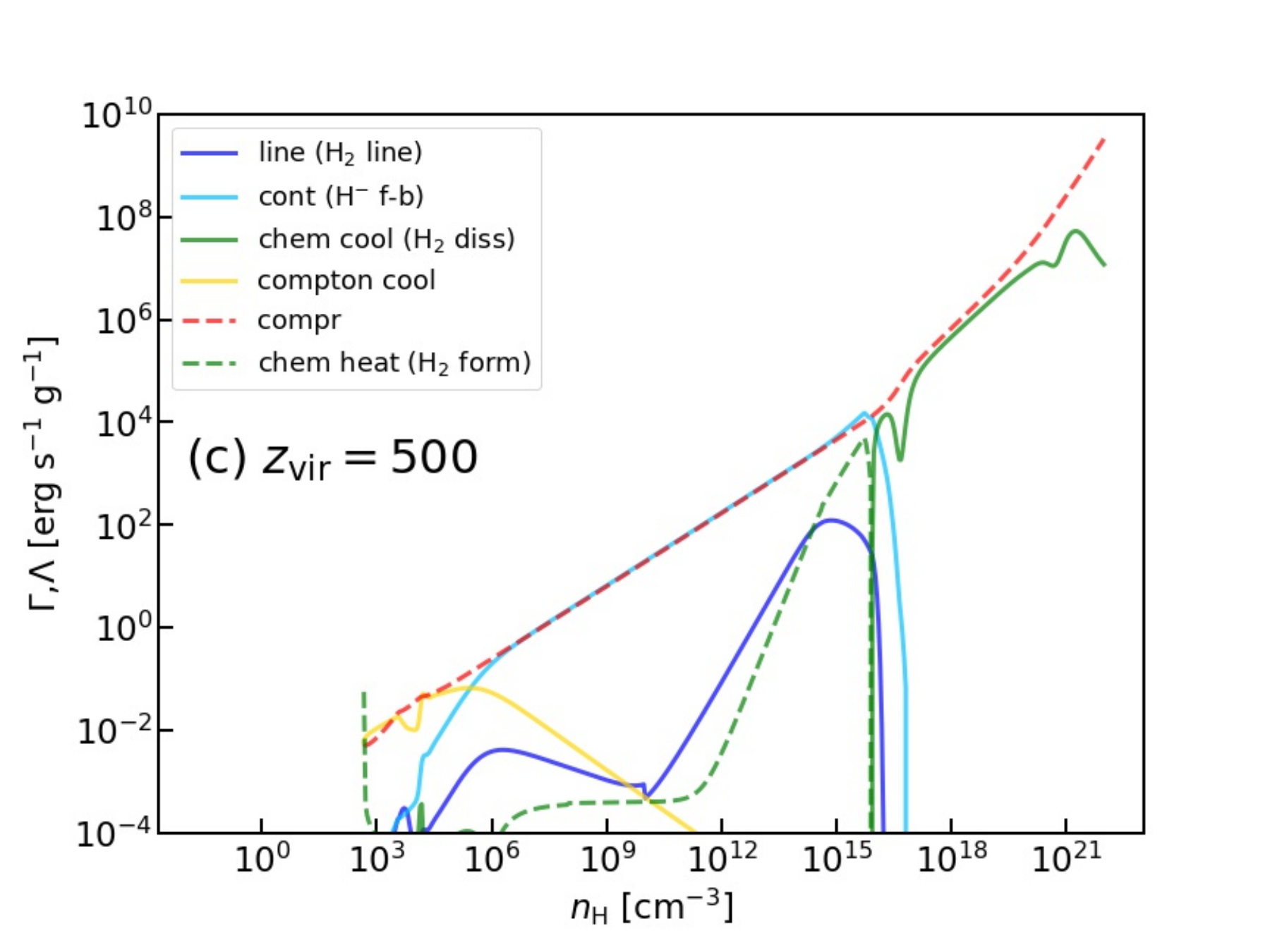}}\\%
    \vspace{2pt}
	\caption{ 
Contributions of cooling and heating processes during protostellar collapse of primordial-gas star-forming clouds at three characteristic epochs, corresponding to (a) $z_{\rm vir}=20$, (b) $z_{\rm vir}=200$, and (c) $z_{\rm vir}=500$. Solid lines represent cooling rates, while dashed lines correspond to heating rates. The legend "line" denotes line cooling, always dominated by H$_2$ in all cases (a)-(c). "cont" represents continuum cooling, driven by H$_2$ collision-induced emission in panels (a) and (b), while by H$^{-}$ free-bound emission in panel (c). "chem cool" indicates cooling from chemical reactions, primarily associated with H$_2$ dissociation. "compton cool" denotes Compton cooling. For heating rates, "compr" stands for compressional heating, and "chem heat" represents heating from chemical reactions, mainly associated with H$_2$ formation. 
In each panel, a single individual process, which is indicated in parentheses in the legend, is almost always dominant among the cooling/heating rates represented by the respective 'line', 'cont', etc.
}
    \label{Fig_cooling}
\end{figure}

Figure \ref{Fig_nT} illustrates the thermal evolution of cloud cores forming at various cosmological epochs, in a much wider range of densities than in Figure \ref{Fig_nT_exp}, extending up to the densities typical of protostar formation, extending up to the densities typical of protostar formation.
For cases with $z_{\rm vir} \leq 400$, where H$_2$ serves as an important coolant, we also present the evolution of H$_2$ abundance and ionization fraction in Figure \ref{Fig_yH2}. Recall that the evolution of the ionization degree is calculated using the RecFast code up to the maximum expansion (represented by the lightly colored portion of the dash-dotted curves in the figure), while the initial H$_2$ abundance is derived from the IGM values provided by \cite{Galli1998}.

In the subsequent discussion, we commence with the familiar case of a first-star forming cloud at a usual epoch of $z_{\rm vir}=20$ and progressively explore the evolution of those forming at earlier epochs, i.e., higher $z_{\rm vir}$.

\subsubsection{Standard collapse sequence for first star formation: $z_{\mathrm{vir}}=20$}

In this section, we revisit the well-known thermal evolution of first-star forming clouds
at a conventional formation epoch, $z_{\mathrm{vir}}=20$, which serves as a benchmark for comparison with those at higher-$z$ cases to be discussed later.

For this case, the temperature evolution is illustrated by the purple line in Figure \ref{Fig_nT}. Additionally, the evolution of hydrogen molecule and electron abundances within the cloud as a function of density is depicted by the purple lines in Figure \ref{Fig_yH2}. Furthermore, Figure \ref{Fig_cooling} (a) displays the contribution of individual processes to the cooling and heating rates at various densities.

Initially, up to $n_{\mathrm{H}} \sim 1$ cm$^{-3}$, the temperature increases adiabatically with density due to the absence of effective radiative cooling. During this phase, compressional heating dominates among the cooling/heating processes (as observed in Figure \ref{Fig_cooling} a). Once the density reaches $n_{\mathrm{H}} \sim 1$ cm$^{-3}$, the temperature exceeds $\gtrsim 1000 {~\rm K}$, facilitating H$_{2}$ formation via the H$^{-}$ channel reaction:
\begin{equation}
    \begin{aligned}
        \text{H} + \text{e}
	&\rightarrow
	\text{H}^{-} + \gamma \\
	\text{H}^{-} + \text{H} 
	&\rightarrow
	\text{H}_{2} + e
    \end{aligned}
\end{equation}
This leads to the conversion of about 1/1000 of hydrogen into molecular form, but the H$_2$ abundance saturates at around this value due to the recombination of electrons, catalysts of the above reaction (Figure \ref{Fig_yH2}). Consequently, the star-forming clouds cool via H$_2$ line emission, causing the temperature to decrease (Figure \ref{Fig_nT}).

As the density increases to $n_{\mathrm{H}} \sim 10^3-10^{4}$ cm$^{-3}$, the temperature reaches a minimum value of 200K, where the rotational level population of hydrogen molecules attains LTE. Beyond this density, radiative cooling becomes less efficient compared to compressional heating, resulting in a gradual temperature rise.

Subsequently, at densities exceeding $n_{\mathrm{H}} > 10^{9}$cm$^{-3}$, another phase of hydrogen molecule formation commences through three-body reactions, eventually transitioning the primordial gas to an almost fully molecular state (Figure \ref{Fig_yH2}). Despite a slight drop in temperature (Figure \ref{Fig_nT}) accompanying this increase in H$_2$ abundance and subsequent cooling, the chemical heating associated with H$_2$ formation and photon trapping counterbalances the cooling effect (Figure \ref{Fig_cooling} a). Consequently, the temperature gradually increases once again.

However, at densities around $n_{\mathrm{H}} \sim 10^{13}-10^{14}$ cm$^{-3}$, continuum cooling via H$_2$ collision-induced emission becomes effective (Figure \ref{Fig_cooling} a), momentarily decreasing the temperature (Figure \ref{Fig_nT}). Yet, this cooling is transient, as densities exceeding $n_{\mathrm{H}} \sim 10^{16}$ cm$^{-3}$ render the cloud cores optically thick to the continuous absorption, severely limiting radiative cooling efficiency. Consequently, the temperature rises adiabatically, triggering hydrogen molecule dissociation. Although chemical cooling temporarily mitigates temperature increase during dissociation, once most hydrogen molecules dissociate, the temperature rises rapidly (Figure \ref{Fig_nT}). Subsequently, with no effective cooling processes at $n_{\mathrm{H}} \gtrsim 10^{20}-10^{21}$ cm$^{-3}$, gas temperature increases almost adiabatically, leading to the formation of a hydrostatic protstar by increasing pressure counterbalancing the gravitational pull.  This nascent star initially possesses very small mass ($\sim 10^{-3} ~\mathrm{M}_{\odot}$) corresponding to the local Jeans mass at its birth, but it subsequently increases as gas accretion from the surrounding medium ensues, ultimately determining the final stellar mass, a discussion of which is reserved for Section 4.

\subsubsection{Diminished H$_2$ cooling due to CMB: 
$z_{\mathrm{vir}} =100-400$}
In the case of $z_{\rm vir}=100$, 
the minimum temperature, which is attained when the H$_2$ level reaches the LTE around $n_{\rm H}=10^3-10^4 {\rm cm^{-3}}$, is approximately 300 K, somewhat higher than the standard value of 200 K observed in the $z_{\rm vir}=20$ case (Figure \ref{Fig_nT}). This difference arises from the inability of radiative cooling to lower the temperature below the CMB temperature, known as the so-called CMB floor (see Figure \ref{Fig_nT_exp}).
Although the density at which the temperature reaches around 1000 K is slightly higher than at $z_{\rm vir}=20$, resulting in a slightly higher onset density for H$_2$ formation via the H$^{-}$ channel, the amount of formed H$_2$ and subsequent thermal and chemical evolution remain largely unchanged compared to the $z_{\rm vir}=20$ case (Figure \ref{Fig_yH2}).

For higher $z_{\rm vir}$, however, not only is the CMB floor higher (Figure \ref{Fig_nT_exp}), but also the amount of formed H$_2$ itself is significantly lower, by an order of magnitude or more, at densities below $n_{\rm H} \lesssim 10^{9}{\rm cm^{-3}}$, before the onset of three-body reactions (Figure \ref{Fig_yH2}). This reduction arises from the destruction of intermediate species H$^{-}$, which forms H$_2$, due to photodetachment by CMB photons, hindering H$_2$ formation (a quantitative discussion of this effect will be provided in $\S$ \ref{H-photodiss} later). Consequently, the efficiency of H$_{2}$ cooling decreases, leading to higher temperatures (Figure \ref{Fig_nT} and Figure \ref{Fig_cooling} b).

For $z_{\mathrm{vir}} = 200-400$, the minimum temperature reaches approximately $10^{3}$ K in all cases (Figure \ref{Fig_nT}). Once the density exceeds $n_{\mathrm{H}} \gtrsim 10^{8}$ cm$^{-3}$, H$_2$ formation via three-body reactions initiates, and the subsequent evolution mirrors that of the standard case of star formation at later epochs. Note also that, for $z_{\rm vir}=300-400$, the temperature exhibits a somewhat abrupt decrease at $n_{\rm H}=10^{8}-10^{10} {\rm cm^{-3}}$ as H$_2$ cooling via three-body reactions begins, owing to the higher temperature before that.

\subsubsection{Direct collapse via atomic cooling: $z_{\mathrm{vir}} > 500$}

The thermal evolution of clouds forming at earlier epochs, i.e., higher $z_{\rm vir}$, differs completely. As evident from Figures \ref{Fig_nT_exp} and \ref{Fig_nT}, once Compton-decoupled from the CMB, the temperature continues to rise, reaching as high as 5000-6000 K. Subsequently, the evolution proceeds nearly isothermally until a density as high as $n_{\mathrm{H}} \sim 10^{16}$ cm$^{-3}$.

This behavior closely resembles the so-called direct collapse scenario, observed at lower $z_{\rm zir} \sim 10$, where H$_2$ formation and its subsequent cooling are inhibited by such processes as far-ultraviolet irradiation \citep{Omukai2001}, dense shock from cold streams \citep{Inayoshi2012, Inayoshi2015, Kiyuna2023}, and dynamical heating from other halo mergers \citep{Wise2019}. In such cases, the collapse occurs solely due to atomic cooling (specifically, H$^{-}$ free-bound emission). 
However, in the current study, the inhibition of H$_2$ formation is attributed to H$^{-}$ photodetachment by the CMB.
It should be noted that in the conventional direct collapse scenario at $z \sim 10$, H Ly$\alpha$ cooling becomes significant in the low-density regime ($n_{\rm H}<10^{5}{\rm cm^{-3}}$) while, in our case, within this density range, Compton cooling becomes crucial, followed by cooling via H$^{-}$ free-bound emission at higher densities (illustrated as "cont" in Fig. \ref{Fig_cooling}c), as in the standard case.

Following this thermal evolution, star-forming clouds experience minimal fragmentation, undergoing monolithic collapse into massive clumps at their centers \citep{Latif2013, Inayoshi2014}. Ultimately, supermassive stars exceeding approximately $10^{5}~\mathrm{M}_{\odot}$ are formed. It is conjectured that these supermassive stars undergo gravitational collapse due to general relativistic instabilities, known as post-Newtonian instability, eventually leading to the formation of black holes \citep{Chandrasekhar1964, Umeda2016, Woods2017, Haemmerle2018}.


\subsection{Quantitative analysis of photo-reactions}

\subsubsection{H$^{-}$, H$_{2}^{+}$ photodissociation}
\label{H-photodiss}
\begin{figure}[h]
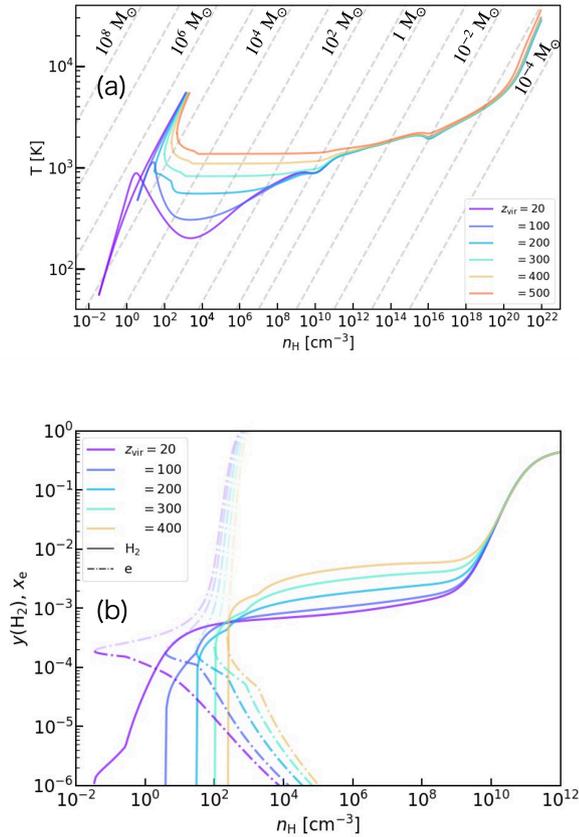

    \subfigure{%
        \includegraphics[keepaspectratio, bb=0 0 6144 3948, width = \columnwidth]{./figures/nT_no_pd.pdf}}
    \hspace{0pt}
    \subfigure{%
        \includegraphics[keepaspectratio, bb=0 0 6144 4608, width = \columnwidth]{./figures/yH2_large_no_pd.pdf}}\\%
    \caption{
    Same as in (a) Figure 2 and (b) Figure 3, respectively, but without considering H$^{-}$ photodetachment.
    }
    \label{Fig_noH-photodiss}
\end{figure}

So far, we have observed that H$_2$ formation reactions 
in star-forming clouds at extremely high redshifts 
are inhibited by CMB photons, resulting in higher temperatures compared to the standard case of $z_{\rm vir}=20$. Now, we quantitatively estimate when such photo-chemical reactions become significant.

First, let us examine the thermal evolution and the H$_{2}$ and $e$ abundances without considering H$^{-}$ photodetachment. These are illustrated in Fig \ref{Fig_noH-photodiss}, showing significant differences from Figs \ref{Fig_nT} and \ref{Fig_yH2}, where H$^{-}$ photodetachment is properly considered. In Fig \ref{Fig_noH-photodiss}(b), molecular hydrogen is abundant with $10^{-3}-10^{-2}$, especially at higher $z$, due to the increased gas temperature. The temperatures reach the CMB floors, leading to isothermal evolution at low densities (Fig \ref{Fig_noH-photodiss} a). Particularly, a significant difference in the H$_{2}$ abundance is observed at $z_{\mathrm{vir}} \gtrsim 200$.
This can be understood from the following analysis.

\begin{figure}[hbtp]
    \centering
	\includegraphics[keepaspectratio, bb=0 0 864 648, width = \columnwidth]{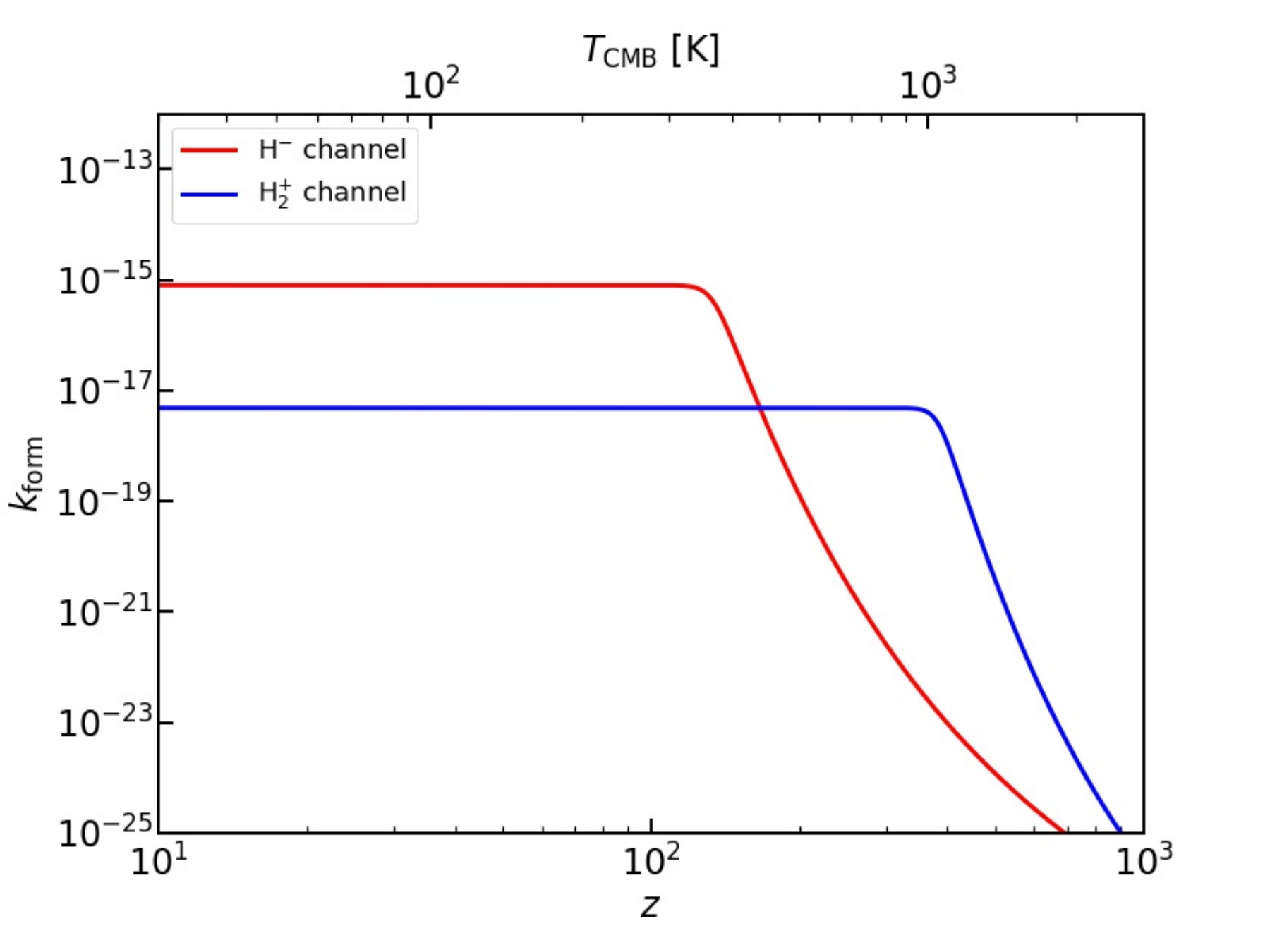}
	\caption{
 Effective reaction coefficients for H$_2$ formation reactions via the H$^-$ channel (red) and the H$_2^{+}$ channel (blue), plotted as a function of redshift. Corresponding radiation temperatures are also indicated on the upper axis. Gas temperature and density are set at representative values of 1000 K and $10^{3} ~\mathrm{cm^{-3}}$, respectively.
 Here, the reaction coefficients for different CMB temperature are approximated using formulas provided by \cite{Galli1998}, which closely match the numerical values obtained in our calculations. For H$_2^{+}$ photodissociation, the LTE expression in \cite{Galli1998} is used.
 }
	\label{Fig_kform}
\end{figure}

At low densities, H$_2$ formation primarily occurs through the H$^{-}$ channel. However, in the presence of radiation fields (above a threshold of $>0.76$ eV), H$^{-}$ ions are photo-detached. The relevant chemical reactions for this channel can be described as follows:
\begin{equation}
\begin{aligned}
\text{H} + \text{e}
&\overset{k_{1}}{\underset{k_{2}}{\rightleftarrows}}
\text{H}^{-} + \gamma \\
\text{H}^{-} + \text{H}
&\underset{k_{3}}{\rightarrow}
\text{H}_{2} + e , 
\end{aligned}
\end{equation}
where associated $k$'s signify respective reaction coefficients.  
The fraction of H$^{-}$ ions formed via reaction 1 and used in H$_2$ formation is $k_3 n_{\rm H}/(k_2+k_3 n_{\rm H})$. Thus, the effective rate of H$_2$ formation is given by
\begin{equation}
k_{\mathrm{form}}=
k_{1} \times \frac{k_{3} n_{\mathrm{H}}}{k_{2} + k_{3} n_{\mathrm{H}}} .
\end{equation}
This effective H$_2$ formation reaction coefficient is plotted in Fig \ref{Fig_kform} (red line). Representative values of gas density and temperature ($n_{\mathrm{H}}=1000$ cm$^{-3}$, $T=1000$ K) are used. From this figure, it is evident that the formation rate via the H$^{-}$ channel significantly decreases under the influence of the CMB for $T_{\rm CMB} \gtrsim 400$ K ($z \gtrsim 130$).

In addition to the H$^{-}$ channel, there exists another H$_2$ formation pathway called the H$_2^{+}$ channel, catalyzed by H$^{+}$ ions, where H$_2^{+}$ acts as an intermediate:
\begin{equation}
\begin{aligned}
\text{H} + \text{H}^{+}
&\overset{k_{1}}{\underset{k_{2}}{\rightleftarrows}}
\text{H}_{2}^{+} + \gamma \\
\text{H}_{2}^{+} + \text{H}
&\underset{k_{3}}{\rightarrow}
\text{H}_{2} + \text{H}^{+} .
\end{aligned}
\end{equation}
In the absence of strong radiation fields, this pathway is less significant compared to the H$^{-}$ channel due to lower formation rates by approximately two orders of magnitude. However, when the H$^{-}$ channel is blocked by radiation fields such as the CMB, the H$_2^{+}$ channel becomes important because of the higher photo-dissociation threshold of H$_2^{+}$ ($2.79$ eV from the ground state). The effective formation rate through this channel is also plotted in Fig \ref{Fig_kform} (blue line). It is evident from this figure that the formation rate via the H$_2^{+}$ channel surpasses that of the H$^{-}$ channel for $z \gtrsim 200$. The H$_2$ formation via the H$_2^{+}$ channel has been confirmed to be effective in star-forming regions at such high-redshift epochs
also by cosmological simulations by \cite{Hirano2015}.

Furthermore, at $z \gtrsim 400$, the formation rate via the H$_2^{+}$ channel also decreases significantly due to photodissociation of H$_2^{+}$. As a result, H$_2$ formation is strongly suppressed at higher redshifts, and gas cooling relies solely on atomic cooling. This is consistent with the findings in Sections 3.1 and 3.2, indicating that the collapse of star-forming clouds during this epoch follows the so-called direct collapse scenario.

\subsection{H$_{2}$ photodissociation}

In addition to the indirect inhibition of H$_2$ formation due to the destruction of intermediate products such as H$^{-}$ and H$_2^{+}$ via their photodetachment, the H$_2$ abundance can also be reduced by photodissociation of H$_2$ itself. Here, we aim to estimate when the latter effect becomes significant.

In a cloud collapsing at the free-fall rate, the H$_2$ formation reaction proceeds as much as possible in the free-fall time $t_{\rm ff}$, producing H$_2$ with its fraction $y(\mathrm{H}_2) \simeq k_{\rm form} n_{\mathrm{H}} x_{\mathrm{e}} t_{\mathrm{ff}}$, where $x_{\mathrm{e}}$ is the ionization degree. In other words, the timescale for molecular hydrogen formation also becomes $t_{\rm ff}$. Estimating this with the virialization density $\rho_{\rm vir}$, we have:
\begin{equation}
t_{\mathrm{ff}}
=
\sqrt{\frac{3 \pi}{32 \mathrm{G} \rho_{\mathrm{vir}}}}
\simeq
2.98 \times 10^{12}
\left( \frac{1081}{1+z} \right)^{3/2} ~[\text{s}].
\end{equation}
On the other hand, the timescale for photodissociation is given by the inverse of the photodissociation rate, $k_{\rm pd}^{-1}$. Hence, the condition for photodissociation to become significant is when $t_{\rm ff}>k_{\rm pd}^{-1}$.

Since the frequency at which photodissociation occurs ($\simeq 12.4$ eV) is greater than the peak frequency of the CMB, using the Wien approximation and neglecting the shielding effect, we have:
\begin{equation}
\frac{1}{k_{\mathrm{pd}}} \simeq 1.8 \times 10^{-9}
\exp \left(\frac{52800}{1+z} \right) [{\rm s}].
\end{equation}

This becomes smaller than $t_{\rm ff}$ when $1+z > 1081$. Furthermore, due to the exponential dependence of the photodissociation rate on $1+z$, it is evident that photodissociation becomes unimportant very rapidly as $z$ decreases. Thus, in the post-recombination era we are considering ($z < 1000$), photodissociation of H$_2$ by the CMB is entirely negligible.

\section{Discussion}

\subsection{Estimating the mass of forming stars}
\begin{figure}[h]
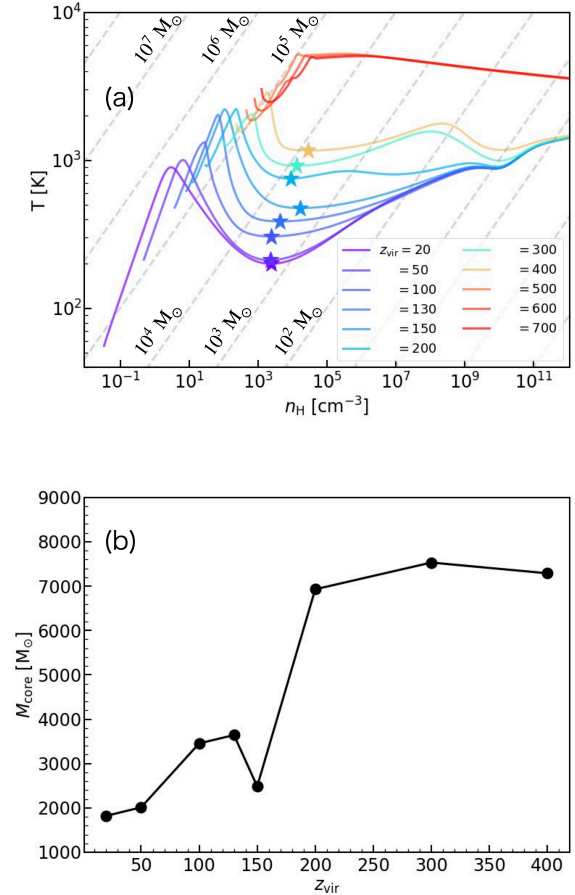

    \subfigure{%
	\includegraphics[keepaspectratio, bb=0 0 6144 4608, width = \columnwidth]{./figures/nT_star_.pdf}}%
    \hspace{0pt}
    \subfigure{%
	\includegraphics[keepaspectratio, bb=0 0 6144 4608, width = \columnwidth]{./figures/Jeansmass.pdf}}\\%
    \caption{
Mass of dense cores formed within primordial-gas star-forming clouds at each virialization epoch.
(a) Temperature evolution as in Figure 2, emphasizing the low-density regime. The Jeans mass (grey dashed line) at the time of minimum temperature (star symbol, denoting the so-called loitering phase) becomes the fragmentation mass, i.e., the mass of the dense core.
(b) Plot of the dense core mass for each $z_{\rm vir}$.
    }
    \label{Fig_nT_star}
\end{figure}
\begin{figure}[hbtp]
    \centering
	\includegraphics[keepaspectratio, bb=0 0 864 648, width = \columnwidth]{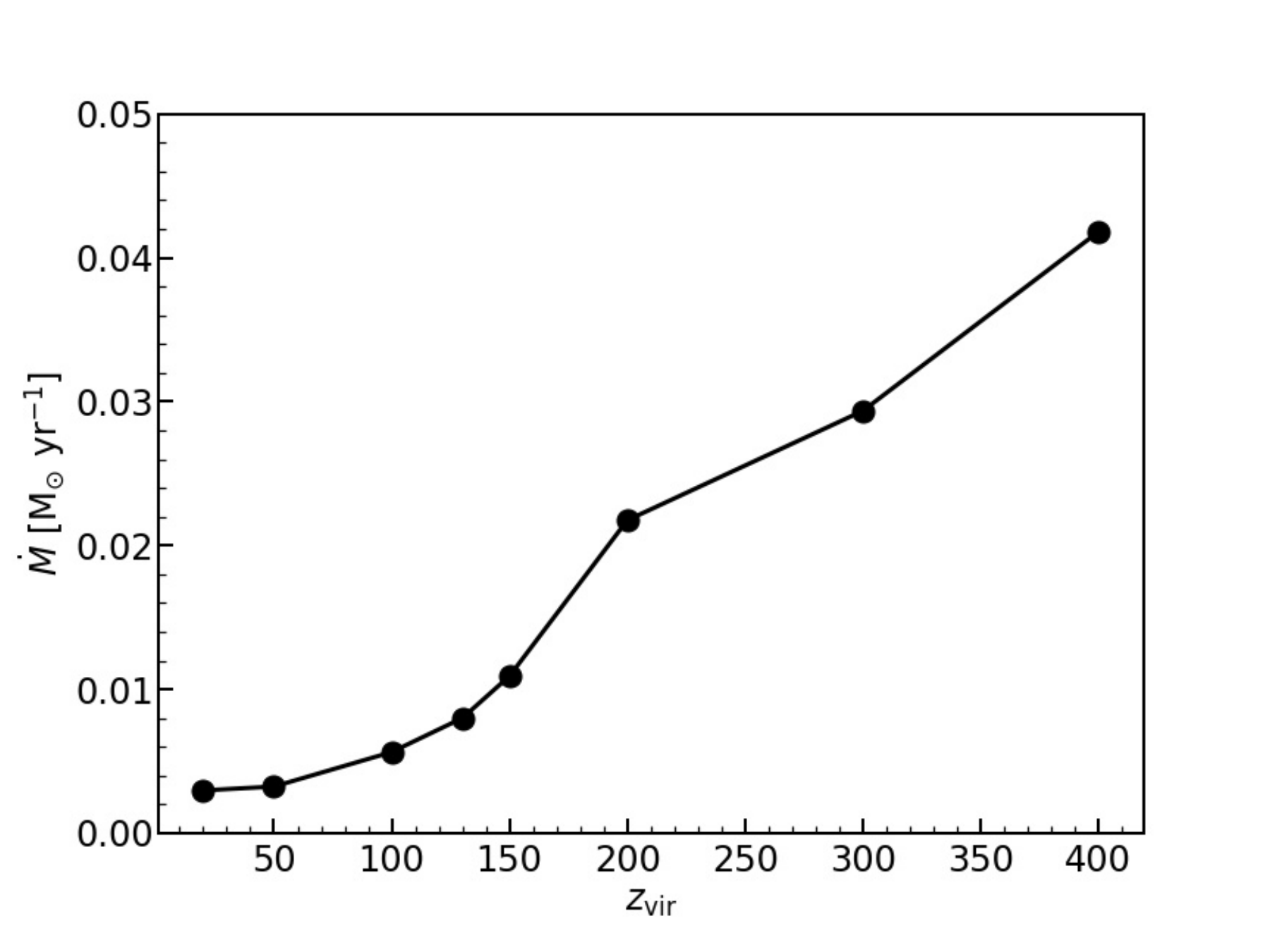}
	\caption{
 Accretion rates on core-scale estimated from the temperature at the time of dense core formation. Plotted for each formation epoch $z_{\rm vir}$. 
 }
	\label{Fig_accretion}
\end{figure}
\begin{figure}[hbtp]
    \centering
	\includegraphics[keepaspectratio, bb=0 0 864 648, width = \columnwidth]{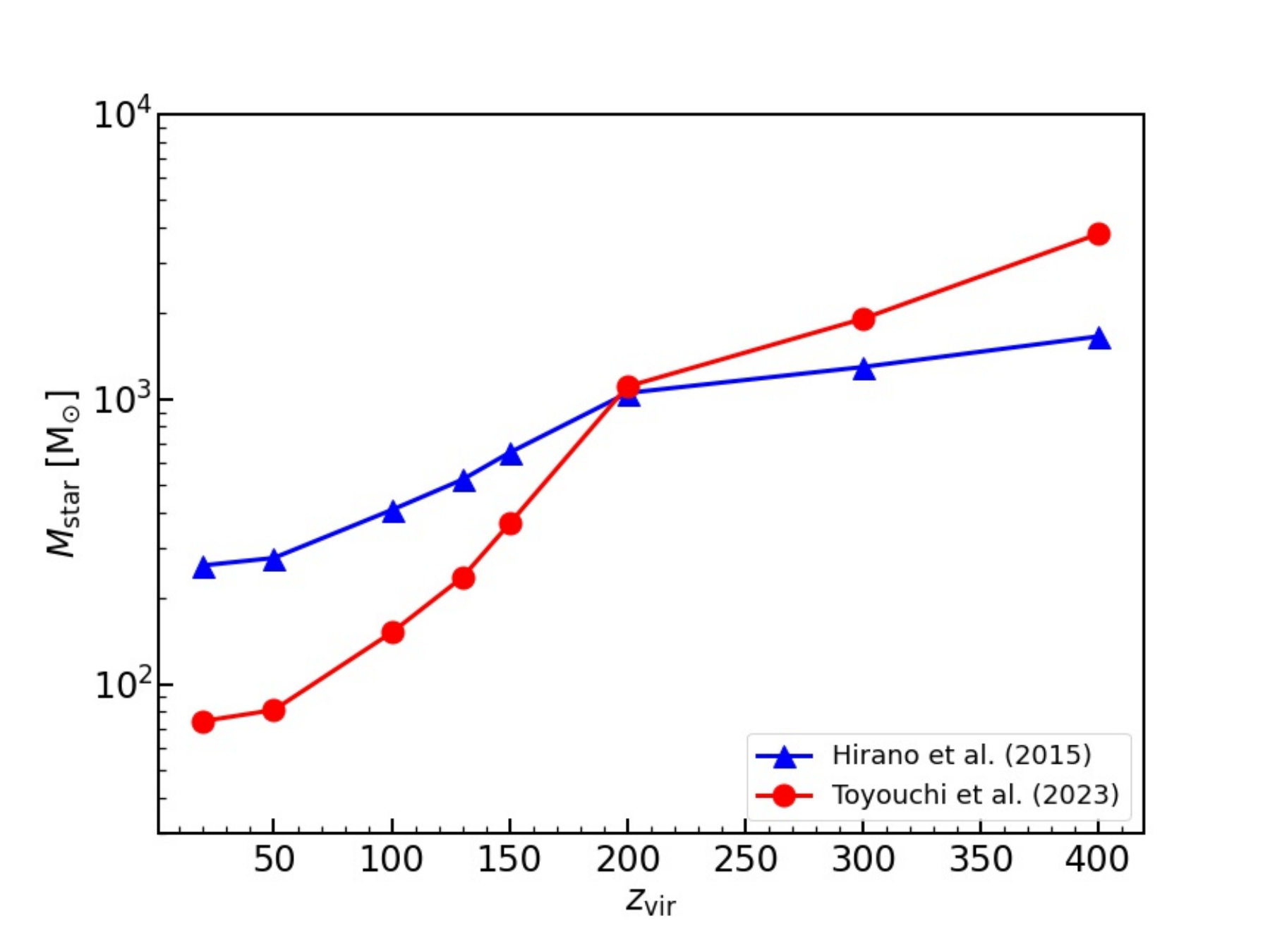}
	\caption{
 Final masses of forming first stars estimated using the relationship between accretion rates and formed stars obtained from previous simulations. Plotted for two different relationships (blue: \cite{Hirano2015}, red: \cite{Toyouchi2023}).
}
	\label{Fig_finalmass}
\end{figure}
Here, by using the temperature evolution of the star-forming clouds obtained in the preceding section, we estimate the masses of stars formed. First, as an upper limit for the mass of stars, we evaluate the mass of dense cores, which serve as the mass reservoir for star formation.
When the temperature in clouds decreases with increasing density while undergoing gravitational collapse, the clouds tend to become filamentary in shape. Later, when the temperature begins to increase, these clouds fragment into clumps that are about the size of the Jeans mass at that moment, leading to the formation of dense cores \citep{Inutsuka1992, Li2003, Jappsen2005}.
Even when fragmentation is not as efficient as envisaged above, a temporary slowdown in contraction occurs due to the pressure force, known as the "loitering phase," when the temperature shifts from decreasing to increasing \citep{Bromm2002}. Following this, a runaway collapse begins as sufficient mass, exceeding the Jeans mass, accumulates in the central region, leading to the formation of a dense core. Therefore, the Jeans mass at the density where temperature is minimized serves as an indicator of the mass of dense cores \citep{Schneider2002, Omukai2005}, as plotted in the figure illustrating the temperature evolution of the star-forming clouds as a function of density (Figure \ref{Fig_nT_star}a). The Jeans mass at this temperature minimum for each formation epoch is plotted in Figure \ref{Fig_nT_star} b. For typical first stars (at $z_{\rm vir}=20$), this Jeans mass is approximately $2000~\mathrm{M}_{\odot}$, consistent with the mass of dense cores obtained from numerical simulations \citep{Bromm2002, Abel2002, Yoshida2006}. 
As redshift increases, the temperature rises while the fragmentation density remains relatively constant. However, at $z=150$, this density is slightly higher due to the subtlety of the almost isothermal temperature evolution, leading to a temporary decrease in the fragmentation mass. Nonetheless, as the redshift exceeds 200, larger dense-core masses are observed, reaching approximately $7000~\mathrm{M}_{\odot}$.
The estimated mass of dense cores is significantly larger than the mass determined by stellar radiative feedback, as estimated below, indicating the abundant mass reservoir for star formation. For the case of direct collapse at $z>500$, there is no significant temperature decrease after virialization. Hence, the gas collapses nearly isothermally without fragmentation, i.e., monolithic collapse. The necessary gas mass for this collapse is given by the maximum Jeans mass during the collapse evolution, exceeding $10^{5}~\mathrm{M}_{\odot}$.

Within these dense cores, after the runaway collapse, the central regions become extremely dense, reaching densities as high as $10^{20}{\rm cm^{-3}}$. At this point, the temperature evolution becomes adiabatic, and gravitational collapse is halted by pressure gradient forces, leading to the formation of protostars in hydrostatic equilibrium \citep{Omukai1998, Yoshida2008, Greif2011}. Initially, the mass of these protostars is very small, but it increases subsequently due to accretion from the surrounding gas.

The accretion rate is related to the temperature, hence the sound speed $c_{\rm s}$, of the dense core and is given by:
\begin{equation}
    \dot{M} \simeq \phi \frac{c_{\rm s}^3}{\mathrm{G}}.
\end{equation}
Here, $\phi$ is a dimensionless number typically ranging from $O(1)$ to $O(10)$, representing the degree to which the collapse toward protostar formation proceeds dynamically \citep{Whitworth1985}. For instance, in the Larson-Penston solution \citep{Larson1969, Penston1969}, which is
the self-similar solution corresponding to the dynamic limit of protostar formation through runaway collapse, $\phi=47$. In contrast, for the Shu solution \citep{Shu1977}, which is the self-similar solution where gas accretes statically from unstable hydrostatic equilibrium cores, $\phi=0.975$. When collapse begins from realistic initial conditions, it is known to take values between these extremes \citep{Foster1993}.

A protostar growing in this manner often have its final mass determined by the radiative feedback from itself, which stops accretion before all the material within the dense core, acting as a mass reservoir, accretes onto the star \citep{Larson1971, Nakano1995}.
This radiation feedback, in the case of first stars, occurs when the stellar mass reaches several tens of $\mathrm{M}_{\odot}$ \citep{McKee2008, Hosokawa2011}. At this point, intense ultraviolet radiation from the star irradiates the circumstellar accretion disk, causing the gas in the disk to evaporate due to ionization heating, thereby halting accretion.

\citet{Hirano2014, Hirano2015a} conducted 2D radiation hydrodynamic simulations of primordial-gas clouds within minihalos, chosen from cosmological simulations as potential sites for first star formation. They investigated the final mass of stars due to radiation feedback during their growth. Their findings revealed a strong correlation between the accretion rate on the scale of dense cores and the mass of stars formed, which can be expressed by the following equation \citep{Hirano2015a}:
\begin{equation}
    M_{\ast} = 250 ~\mathrm{M}_{\odot} \left(\frac{\dot{M}}{2.8 \times 10^{-3} ~\mathrm{M}_{\odot} ~{\rm yr^{-1}}} \right)^{0.7}.
\end{equation} 
On the other hand, \citet{Toyouchi2023} performed similar calculations for more massive atomic cooling halos, which are expected to give rise to more massive stars. They found a relationship between mass and accretion rate with even stronger dependence. Their numerical results were consistent with an analytical model of photoevaporation of circumstellar disks \citep{Tanaka2013}. This relationship, although not expressible by a simple power-law, can be understood by referring to their Figure 10.

In our calculations, we estimate the accretion rate on the scale of dense cores based on their temperature, i.e., the minimum temperature. In doing so, we chose the dimensionless number $\phi$ to be 8.06 to ensure that the accretion rate estimated from the typical temperature evolution during first star formation in our calculations (at $z=20$) matches the typical value of the accretion rate reported by \citet{Hirano2015a}, which is $3\times 10^{-3} ~\mathrm{M}_{\odot} ~{\rm yr^{-1}}$.
The typical accretion rate obtained using this approach for each redshift is depicted in Figure \ref{Fig_accretion}. Accretion rates increase with higher redshifts, reflecting higher core temperatures. For instance, at redshifts below 100, typical accretion rates match the value of $3\times 10^{-3} ~\mathrm{M}_{\odot} ~{\rm yr^{-1}}$ for standard first star formation. However, at redshifts ranging from 200 to 400, they increase to approximately $2-4 \times 10^{-2} ~\mathrm{M}_{\odot} ~{\rm yr^{-1}}$, roughly an order of magnitude higher.

The estimated masses of stars based on these accretion rates are depicted in Figure \ref{Fig_finalmass}, with plots corresponding to the relationships proposed by \citet{Hirano2015a} and \citet{Toyouchi2023}. Since our case encompasses various cases from molecular to atomic cooling, it is not entirely clear which relationship is more appropriate. Therefore, we discuss each relationship separately.
If we adopt the relationship proposed by \citet{Hirano2015a}, the mass of stars ranges from $260 ~\mathrm{M}_{\odot}$ at $z=20$ to increasing masses at earlier cosmic epochs, exceeding $1000~\mathrm{M}_{\odot}$ for $z>200$ and reaching $1660~\mathrm{M}_{\odot}$ at $z=400$. On the other hand, \citet{Toyouchi2023}'s relationship shows a stronger dependence of mass on accretion rate. At $z=20$, the mass is $74~\mathrm{M}_{\odot}$, but it rapidly increases to over $1000~\mathrm{M}_{\odot}$ at $z=200$ and reaches $3800~\mathrm{M}_{\odot}$ at $z=400$.
Regardless of which relationship is employed, the formation of very massive stars, approximately $1000~\mathrm{M}_{\odot}$ in mass, occurs for $z>200$. Moreover, in cases where $z>500$, gigantic clouds undergo direct collapse, giving rise to supermassive stars. In these instances, protostars formed at the core of the clouds rapidly become supermassive at tremendous accretion rates of $0.1-1 ~\mathrm{M}_{\odot} ~{\rm yr^{-1}}$. They quickly reach masses of approximately $10^{5}~\mathrm{M}_{\odot}$ before collapsing into black holes due to general relativistic instabilities.

\subsection{Caveats}

\subsubsection{Impact of Compton drag on gas dynamics}
In our calculations, we have assumed nearly free-fall, runaway collapse for the dynamics of star-forming clouds. However, in the early universe where the radiation energy is high, consideration must be given to radiative viscosity, commonly known as Compton drag \citep{Umemura1993}. The Compton drag force acting on unit mass of gas moving with peculiar velocity $v$ relative to the cosmic expansion is given by:
\begin{equation}
f_{\rm drag} = \frac{4}{3}
\frac{\sigma_{\rm T} a T_{\rm CMB}^4 n_{\rm e}} {\rho c} v
\simeq \frac{\sigma_{\rm T} a T_{\rm CMB}^4 x_{\rm e} }{ m_{\rm H} c} v.
\end{equation}
Here, $\sigma_{\text{T}}$ is the Thomson scattering cross-section, $a$ the radiation energy constant, $n_{\text{e}}$ the electron number density, and $x_{\text{e}}$ the ionization degree. Consequently, the timescale on which drag influences gas motion is given by:
\begin{equation}
t_{\rm drag} = \frac{m_{\rm H} c}{\sigma_{\rm T} a T_{\rm CMB}^4 x_{\rm e}}.
\end{equation}
Comparing this timescale with the timescale of cosmic expansion, characterized by the Hubble time:
\begin{equation}
t_{\rm H} = H_{0}^{-1} \Omega_{m}^{-1/2} (1+z)^{-3/2},
\end{equation}
we find that the era when Compton drag becomes significant:
\begin{equation}
 1+z > 140 x_{\rm e}^{-2/5}.     
\end{equation}
This indicates the necessity of considering the drag force for ionized gas at $z \gtrsim 100-200$. On the other hand, the star-forming clouds under consideration are predominantly neutral after the cosmic recombination, with a low ionization fraction of approximately $x_{\text{e}} \simeq 10^{-4}$. Thus, drag can be neglected, justifying our assumptions.

\subsubsection{Impact of baryon streaming motions}
Supersonic coherent baryonic flows, known as streaming motions, exist just after cosmic recombination. These flows have root-mean-square velocities of $\sigma_{\rm rms} \simeq 30 , \text{km/s}$ on scales of a few comoving Mpc \citep{Tseliakhovich2010}. The relative streaming motions between dark matter and baryons can lead to a suppression of halo formation in the early universe and a reduction in the gas content within these haloes, thereby increasing the minimum mass required for cooling and star formation \citep{Tseliakhovich2011, Stacy2011, Maio2011, Schauer2019, Schauer2021}. This effect significantly delays the formation of the very first stars in the universe at $z \simeq 60-70$, with an estimated delay time of up to 3 Myrs in standard cosmology \citep{Fialkov2012}.

By analyzing the results of previous numerical simulations, \cite{Fialkov2012} proposed the following equation to determine the minimum cooling threshold for halos in terms of the circular velocity:
\begin{equation}
V_{\text{cool}}(z) = \sqrt{(3.714 , \text{km/s})^2 + [4.015 \cdot v_{\text{bc}}(z)]^2}.
\label{eq:Vcool}
\end{equation}
In this equation, the first term under the square root on the right-hand side represents the threshold for molecular cooling, while the second term accounts for the effect of baryon streaming motions. It is important to note that the streaming motions are decelerated by cosmic expansion, as described by:
\begin{equation}
v_{\text{bc}}(z)= v_{\text{bc,rec}} \left( \frac{1+z}{1100} \right),
\end{equation}
where $v_{\text{bc,rec}}$ is the streaming velocity at the time of recombination ($z=1100$).

Below we estimate the influence of streaming velocities on the formation of halos and the stars within them. Using the most common velocity in the distribution, $v_{\text{bc,rec}} = 0.82 \sigma_{\rm rms}$ 
and comparing the first and second terms under the square root in equation \eqref{eq:Vcool}, we observe that the effect of streaming velocity becomes significant at redshifts
$z > 40$. 
At the very early epochs $z>100$ we are considering, 
the first term can be neglected:
\begin{equation}
V_{\text{cool}}(z) = 8.7 , \text{km/s} \left(\frac{1+z}{100}\right)
\end{equation}
for the most common velocity. The corresponding halo mass is given by:
\begin{equation}
M_{\rm cool} = \frac{r_{\rm vir} V_{\rm cool}^2}{G}.
\end{equation}
Substituting equation \eqref{eq:rvir} for $r_{\rm vir}$
we find
\begin{equation}
M_{\rm cool} = 4 \times 10^5 M_{\odot} \left(\frac{1+z}{100}\right)^{3/2}.
\end{equation}
The effects of baryon streaming motions are negligible for halos more massive than this threshold. Since the halos we are considering have virial masses $M_{\rm vir} > 10^{6}-10^{7} M_{\odot}$ as mentioned in Section 3.1, we expect that streaming motions will not significantly influence the collapse of these halos, except perhaps at the highest redshifts we have considered.

Nevertheless, while gravitational collapse of these clouds leads to star formation despite streaming motions, the dynamics during collapse may still be affected. In regions of high velocity, the fraction of gas remaining in halos might not only decrease, but even if contraction begins, the gas motion could be heavily influenced by streaming motions. Depending on the velocity, this could either promote the fragmentation of gas clouds or inhibit it, leading to contraction in larger clumps. This complex dependency has been demonstrated in numerical simulations by \cite{Hirano2017} for first-star-forming halos forming at a redshift of $z \simeq 30$.
Investigating how the dynamics of first-star-forming gas behave in response to streaming velocity in the earlier universe remains an intriguing topic for future studies.

\subsection{Future observational prospects}
From our calculation, first stars in the very early universe ($z \gtrsim 100$) typically have masses exceeding several $100~\mathrm{M}_{\odot}$. Such stars wold collapse without explosion, leaving behind intermediate-mass black holes (IMBHs) \citep{Shibata2002, Uchida2017}
although, within a narrow mass range of approximately $5\times 10^4 ~\mathrm{M}_{\odot}$, the collapse due to general relativistic instabilities triggers runaway nuclear reactions leading to explosions \citep{Chen2014, Nagele2022}.

On the other hand, in cases where there is a mass distribution among forming stars, those with somewhat smaller masses ranging from $140-260~\mathrm{M}_{\odot}$ may also form and subsequently undergo pair-instability supernovae (PISNe) \citep{Heger2002}. These supernovae, approximately an order of magnitude more energetic than typical core-collapse supernovae, are anticipated to be detectable up to approximately $z\sim 10$ in forthcoming observations facilitated by instruments such as the Roman Space Telescope or GREX-PLUS \citep{Moriya2022}. However, given the extreme dimness and far-infrared wavelengths associated with stars at $z\sim 100$, direct detection in the near future, even if these stars exist, would likely prove impossible.

Could detection of first stars in the extreme early universe be feasible through some means? Typically, first stars are often formed as binaries 
\citep{Machida2008, Stacy2010, Sugimura2020, Sugimura2023}. Moreover, it is common for supermassive stars to also form as binaries \citep{Chon2020}. After the binary stellar evolution, binary BHs can ultimately form, and if they are sufficiently close ($<0.1$au), they can merge within the age of the universe, making such events observable through gravitational waves.

To estimate when the merger events occur, we need to consider the delay time. This refers to the time taken from the birth of stars to the merger of binary black holes. It is determined by the (very uncertain) separation distance between binaries, which necessitates further research in this area.
If the merger occurs around $z\sim$ several tens, gravitational waves from these IMBH binary mergers can be detected by Laser Interferometer Space Antenna (LISA).
Furthermore, if mergers occur even earlier, the realization of a space experiment using cold atoms to search for ultra-light dark matter and to detect gravitational waves, known as the Atomic Experiment for Dark Matter and Gravity Exploration (AEDGE), could make it possible to observe IMBH mergers even at $z\sim 1000$ \citep{El-Neaj2020}.

If such IMBH merger events were observed in the early universe, they could also potentially originate from primordial black hole (PBH) binaries. While PBHs can form binaries through their three-body interactions, it is generally believed that their binary fraction is lower compared to those originating from first stars \citep{Ioka1998}. Additionally, the redshift distribution of merger events would differ from those originating from first stars, as stellar BH mergers would occur after a certain period of star formation, leading to distinct characteristics \citep[see e.g., ][]{Tanikawa2021, Tanikawa2022}.
Detailed modeling of these scenarios through future study is highly desirable. If IMBHs indeed existed during those periods, it could provide a direct explanation for the presence of supermassive black holes (SMBHs) found at high redshifts ($z > 6-7$) \citep{Mortlock2011, Banados2018, Matsuoka2019}.

In this study, we explore scenarios where the first stars form at significantly earlier epochs (e.g., $z > 100$) than conventionally predicted (typically $z = 20-30$) due to the possible enhancement of density fluctuations on small ($<$Mpc) scales. This is motivated by recent JWST observations revealing an unusual abundance of bright galaxies at high redshifts. Specifically, we examine how the formation processes of these first stars might differ from typical trajectories and assess the masses of the stars that result.

Although we have not adopted a specific model for small-scale deviations in the fluctuation spectrum, estimating the number and timing of first star formation necessitates assuming some specific models. As discussed in Section 1, various models are possible, each featuring characteristics such as a blue-tilt, bumps, a running spectral index etc. in the density fluctuation spectrum on small scales. Integrating these models with our findings on the masses of first stars formed at different epochs will be crucial for developing theoretical predictions that meet observational constraints.
Particularly, identifying which models of the small-scale fluctuation spectrum can explain the observed number density of high-z galaxies by JWST, while also meeting various observational constraints, such as the electron scattering optical depth measured by CMB polarization \citep[e.g.,][]{Visbal2015},  the abundance of dwarf galaxies\citep[e.g.][]{Nakama2018}, and  
the pulsar timing data
\citep[e.g.,][]{Aslanyan2016}, remains a fascinating and challenging task for future research.
\section{Summary}
We have investigated the formation of first stars during the extremely early universe, $z \gtrsim 100$. Employing a one-zone thermochemical model, we have studied the pre-stellar collapse of primordial gas clouds across various formation epochs ranging from a usual epoch $z=20$ to an extremely early epoch of $700$. Our analysis take into accout the influence of the cosmic microwave background (CMB) on radiative heating, Compton cooling, and photodissociation reactions.

We have observed that the influence of the CMB on the evolutionary process is minimal at $z \lesssim 130$, where temperatures closely resemble conventional expectations. However, between $130 \lesssim z \lesssim 500$, H$_{2}$ formation via the H$^{-}$ channel is impeded by CMB-induced photodetachment of H$^{-}$, resulting in higher temperatures compared to the standard thermal evolution. Additionally, at $z \gtrsim 500$, the temperature evolution approaches near-isothermal one around several thousand Kelvins, driven solely by atomic cooling, as the less efficient H$_2^{+}$ channel is also blocked by H$_2^{+}$ photodissociation and thus H$_2$ cooling is entirely suppressed.

Furthermore, we have estimated the mass of forming stars by computing the fragmentation mass and mass accretion rate during the loitering phase. By correlating stellar mass with accretion rate as proposed by 
\cite{Hirano2015a}, we found that for $z<200$, stellar masses range from $70-700 ~\mathrm{M}{\odot}$. At $z>200$, stellar masses exceed $1000 ~\mathrm{M}{\odot}$. For $z>500$, primordial-gas clouds undergo direct collapse, giving rise to supermassive stars more massive than $\sim 10^{5} ~\mathrm{M}_{\odot}$. We conclude that first stars formed at higher redshifts tend to be more massive.

Furthermore, we have estimated the mass of forming stars by computing the fragmentation mass and mass accretion rate during the loitering phase. 
By linking the stellar mass with accretion rates as proposed by \cite{Hirano2015a}, we established that for redshifts $z<200$, stellar masses typically range from 70 to 700 $M_{\odot}$. Remarkably, at $z>200$, the stellar masses exceed 1000$M_{\odot}$, and for $z>500$, primordial-gas clouds undergo direct collapse, forming supermassive stars with masses exceeding $10^{5} M_{\odot}$.
This progression highlights a clear trend: the initial masses of the first stars tend to increase with redshift. The existence of such massive stars in the early universe could potentially be constrained in the future by observations, such as gravitational wave detections from remnant IMBH mergers. Furthermore, the dependency of stellar mass on the formation epoch found in this study will enable us to compute the abundance and timing of the first stars within early galaxies based on specific small-scale fluctuation spectral models. Identifying models that satisfy various observational constraints will be a crucial challenge for future research.
%
%
%

\begin{ack}
We would like to thank K. Eric Sadanari and Shingo Hirano for their helpful comments.
We also would like to thank the anonymous reviewer for the constructive comments, which were instrumental in improving the manuscript.
This work is financially supported by the Grants-in-Aid for Basic Research by the Ministry of Education, Science and Culture of Japan (KO:22H00149). 
\end{ack}


\bibliographystyle{for_pasj}
\bibliography{biblio}


\end{document}